\documentclass[%
 reprint,
showpacs,preprintnumbers,
 amsmath,amssymb,
 aps,
prb,
]{revtex4-1}

\usepackage{graphicx}
\usepackage{dcolumn}
\usepackage{bm}
\usepackage{color}
\usepackage{ulem}

\begin{document}

\preprint{APS/123-QED}

\title{Engineering chiral density waves and topological band structures by \\ 
multiple-$Q$ superpositions of collinear up-up-down-down orders}

\author{Satoru Hayami,$^{1}$ Ryo Ozawa,$^2$ and Yukitoshi Motome$^2$}
\affiliation{
$^1$Department of Physics, Hokkaido University, Sapporo 060-0810, Japan\\
$^2$Department of Applied Physics, University of Tokyo, Tokyo 113-8656, Japan
 }

\begin{abstract}
Magnetic orders characterized by multiple ordering vectors harbor noncollinear and noncoplanar spin textures and can be a source of unusual electronic properties through the spin Berry phase mechanism. 
We theoretically show that such multiple-$Q$ states are stabilized in itinerant magnets in the form of superpositions of collinear up-up-down-down (UUDD) spin states, which accompany the density waves of vector and scalar chirality. 
The result is drawn by examining the ground state of the Kondo lattice model with classical localized moments, especially when the Fermi surface is tuned to be partially nested by the symmetry-related commensurate vectors. 
We unveil the instability toward the multiple-$Q$ UUDD states with chirality density waves, using the perturbative theory, variational calculations, and large-scale Langevin dynamics simulations. 
We also show that the chirality density waves can induce rich nontrivial topology of electronic structures, such as the massless Dirac semimetal, Chern insulator with quantized topological Hall response, and peculiar edge states which depend on the phase of chirality density waves at the edges. 
\end{abstract}
\maketitle

\section{Introduction}
Noncollinear and noncoplanar magnetic orderings have attracted much interest in condensed matter physics, as they often lead to intriguing phenomena and topologically nontrivial electronic states. 
These orders can simultaneously activate secondary order parameters, in addition to the primary magnetic ordering. 
For instance, a noncollinear magnetic order 
carries the vector chirality, which is defined by a vector product of spins, $\langle \mathbf{S}_i \times \mathbf{S}_j \rangle$, while a noncoplanar magnetic order the scalar chirality, which is represented by a triple scalar product, $\langle \mathbf{S}_i \cdot (\mathbf{S}_j \times \mathbf{S}_k) \rangle$. 
Such chirality degrees of freedom generate an emergent electromagnetic field for electrons through the spin Berry phase mechanism, and hence, have a huge potential to induce and control novel electronic states and transport phenomena, such as the anomalous (topological) Hall effect~\cite{Loss_PhysRevB.45.13544,Ye_PhysRevLett.83.3737,Ohgushi_PhysRevB.62.R6065}, orbital and spin current~\cite{Katsura_PhysRevLett.95.057205,Bulaevskii_PhysRevB.78.024402,TaguchiPhysRevB.79.054423}, and magnetoelectric effect~\cite{Bulaevskii_PhysRevB.78.024402}. 
Exploring such unusual states with chirality degrees of freedom is expected to bring a major advance in the field of magnetism and stimulate further possibilities for ``chiraltronics". 

Among them, several noncoplanar magnetic orders have been examined by focusing on the emergence of the anomalous Hall effect. 
Skyrmion crystals are one of the most prominent examples, where the relativistic spin-orbit coupling plays an important role~\cite{Muhlbauer_2009skyrmion,yu2010real,Yu2011,nagaosa2013topological}. 
Another examples have been extensively discussed for 3$d$- and 4$f$-electron compounds, on the basis of the Kondo-type spin-charge coupling on several lattice structures: triangular~\cite{Martin_PhysRevLett.101.156402,Akagi_JPSJ.79.083711,Kumar_PhysRevLett.105.216405,Kato_PhysRevLett.105.266405,Venderbos_PhysRevLett.108.126405,Venderbos_PhysRevB.93.115108}, 
honeycomb~\cite{Jiang_PhysRevLett.114.216402,Venderbos_PhysRevB.93.115108}, kagome~\cite{Ohgushi_PhysRevB.62.R6065,Barros_PhysRevB.90.245119,Ghosh_PhysRevB.93.024401}, square~\cite{Agterberg_PhysRevB.62.13816,Yi_PhysRevB.80.054416}, cubic~\cite{Hayami_PhysRevB.89.085124}, face-centered-cubic~\cite{Shindou_PhysRevLett.87.116801}, and pyrochlore lattices~\cite{Chern:PhysRevLett.105.226403}. 
In particular, the noncoplanar magnetic orders with ferroic ordering of the scalar chirality have attracted much interest, as the spatially uniform scalar chirality generates a coherent spin Berry phase for itinerant electrons and may lead to a quantized anomalous Hall effect~\cite{Ohgushi_PhysRevB.62.R6065,Martin_PhysRevLett.101.156402,Akagi_JPSJ.79.083711,Shindou_PhysRevLett.87.116801}. 
On the other hand, the magnetic states with stripy patterns of the scalar chirality have recently been proposed~\cite{Solenov_PhysRevLett.108.096403,ozawa2015vortex}. 
In these states, the value of the scalar chirality is modulated in real space, and even canceled out in the whole system (the net chirality is zero). 
Thus, these states are regarded as antiferroic-type scalar chirality orderings. 
Given the variety, it is a natural question what is essential for noncollinear and noncoplanar orderings with the chirality degrees of freedom and what determines the spatial pattern of chirality density waves (ChDW). 
It will also be interesting to ask how the different ChDW affect the electronic properties, in both bulk and nanoscale structures, such as topological nature of the band structure, edge states, domain walls, and local electric/spin currents. 

In this paper, we present a systematic theoretical study of vector and scalar ChDW in itinerant electron systems. 
The key ingredient in our study is an up-up-down-down (UUDD) collinear magnetic order [see Fig.~\ref{Fig:multipleQ_ponti}(b)]. 
We demonstrate that a variety of ChDW can be constructed by superpositions of such UUDD orders, which we call multiple-$Q$ UUDD states. 
We examine the instability toward such multiple-$Q$ UUDD states in a minimal model for itinerant magnets, the Kondo lattice model with classical localized moments in two dimensions, using an analytical perturbative expansion with respect to the exchange coupling between itinerant electron spins and localized moments. 
We find that, at particular fillings where the different portions of the Fermi surface are connected by commensurate vectors, the system is unstable toward the multiple-$Q$ UUDD states. 
The higher-order contributions beyond the second-order Ruderman-Kittel-Kasuya-Yosida (RKKY) interaction~\cite{Ruderman,*Kasuya,*Yosida1957} play a key role in this instability. 
While similar mechanisms were discussed for other noncoplanar states~\cite{Solenov_PhysRevLett.108.096403,Akagi_PhysRevLett.108.096401,Hayami_PhysRevB.90.060402,ozawa2015vortex}, our construction has the advantage of extending the variety of ChDW patterns to superstructures beyond one-dimensional stripy ones. 
We carefully confirm the perturbative arguments by two numerical calculations: 
variational calculations for several candidates of the ground state and large-scale Langevin dynamics simulations enabled by the kernel polynomial method (KPM-LD)~\cite{Barros_PhysRevB.88.235101}. 
Furthermore, we find that the ChDW may bring about a topologically nontrivial nature in the itinerant electrons, reflecting oscillations of chirality in real space. 
We show that the system becomes a Dirac semimetal and magnetic Chern insulator depending on the chirality superstructures. 
We also reveal that peculiar edge states appear in these topological states, in different forms depending on where the ChDW are terminated at the edges. 

The rest of the paper is organized as follows. 
In Sec.~\ref{sec:Multiple-$Q$ UUDD State}, after introducing the Kondo lattice model, we briefly discuss how the RKKY interaction fails to determine the ground state of the Kondo lattice model in some particular situations. 
As the candidate for the ground state, we propose multiple-$Q$ modulations of specific UUDD spin states, which result in ChDW. 
In Sec.~\ref{sec:Instability toward Multiple-$Q$ UUDD}, we examine the instability toward the multiple-$Q$ UUDD states by combining the perturbative expansion with respect to the exchange coupling between itinerant and localized spins, variational calculations by using the direct diagonalization of the full Hamiltonian, and the KPM-LD simulations for large-size clusters. 
In Sec.~\ref{sec:Electronic Structure}, we discuss the electronic structure of the multiple-$Q$ UUDD states, with emphasis on the topological properties of bulk band structure and the edge states induced by ChDW. 
We summarize our results in Sec.~\ref{sec:Conclusion}, with making some remarks on the comparison between our multiple-$Q$ UUDD states and other ChDW states.

\section{Multiple-$Q$ UUDD State}
\label{sec:Multiple-$Q$ UUDD State}
In this section, we present the fundamental concept of the multiple-$Q$ UUDD states with ChDW, whose stability is examined in the later sections. 
First, we introduce the Kondo lattice model consisting of classical localized spins and itinerant electrons in Sec.~\ref{sec:Kondo Lattice Model}. 
Then, in Sec.~\ref{sec:RKKY Interaction}, we briefly review the RKKY interaction, which is an effective magnetic interaction appearing in the second-order perturbation with respect to the exchange coupling in the Kondo lattice model. 
After presenting the magnetic structure for the single-$Q$ (1$Q$) UUDD state in Sec.~\ref{sec:UUDD Magnetic Structure}, we describe how to construct the multiple-$Q$ UUDD states on the square and triangular lattices in Sec.~\ref{sec:Spin and Chirality Configurations for Multiple-$Q$ UUDD States}. 
We show that the multiple-$Q$ UUDD states are energetically degenerate with the 1$Q$ one at the level of the RKKY interaction. 
In Sec.~\ref{sec:Chirality Density Waves}, we show the multiple-$Q$ UUDD states possess the real-space superstructures of vector and scalar chirality (ChDW). 

\subsection{Kondo Lattice Model}
\label{sec:Kondo Lattice Model}
We consider a model consisting of noninteracting electrons coupled with localized spins, called the Kondo lattice model, on the square and triangular lattices. 
The Hamiltonian is given by 
\begin{align}
\label{eq:Ham}
\mathcal{H} = &-t_1 \sum_{\langle i, j \rangle \sigma} (c^{\dagger}_{i\sigma}c_{j \sigma}+ {\rm H.c.})
-t_2 \sum_{\langle \langle i, j \rangle \rangle \sigma} (c^{\dagger}_{i\sigma}c_{j \sigma}+ {\rm H.c.}) \nonumber \\
&+J \sum_{i \sigma \sigma'} c^{\dagger}_{i\sigma} \bm{\sigma}_{\sigma \sigma'} c_{i \sigma'}
\cdot \mathbf{S}_i -\mu\sum_{i\sigma}c^{\dagger}_{i\sigma}c_{i\sigma}, 
\end{align}
where $c^{\dagger}_{i\sigma}$ ($c_{i \sigma}$) is a creation (annihilation) operator of an itinerant electron at site $i$ and spin $\sigma$, $\bm{\sigma}=(\sigma^x,\sigma^y,\sigma^z)$ is the vector of Pauli matrices, $\mathbf{S}_i$ is a classical localized spin at site $i$ whose amplitude is normalized as $|\mathbf{S}_i|=1$, $J$ is the exchange coupling constant (the sign is irrelevant for classical localized spins), and $\mu$ is the chemical potential. 
The sums of $\langle i,j \rangle$ and $\langle \langle  i,j \rangle \rangle$ are taken over the nearest-neighbor and second-neighbor sites, respectively, on the square and triangular lattices. 
Hereafter, we take $t_1=1$ as an energy unit. 

In the wave number representation, the Hamiltonian in Eq.~(\ref{eq:Ham}) is transformed into  
\begin{align}
\label{eq:Ham_krep}
\mathcal{H}=\sum_{\mathbf{k} \sigma} (\varepsilon_\mathbf{k}-\mu) c^{\dagger}_{\mathbf{k}\sigma}c_{\mathbf{k}\sigma} + J \sum_{\mathbf{k}\mathbf{q}\sigma\sigma'}
c^{\dagger}_{\mathbf{k}\sigma}\bm{\sigma}_{\sigma \sigma'}c_{\mathbf{k}+\mathbf{q}\sigma'} \cdot \mathbf{S}_{\mathbf{q}}, 
\end{align}
where $c^{\dagger}_{\mathbf{k} \sigma}$ and $c_{\mathbf{k} \sigma}$ are the Fourier transform of $c^{\dagger}_{i\sigma}$ and $c_{i \sigma}$, respectively. 
$\mathbf{S}_{\mathbf{q}}$ is the Fourier transform of $\mathbf{S}_{i}$. 
In Eq.~(\ref{eq:Ham_krep}), $\varepsilon_{\mathbf{k}}$ is the free electron dispersion, which is given by 
\begin{align}
\varepsilon_{\mathbf{k}} = -2\sum_{l=1, 2}(t_1 \cos \mathbf{k}\cdot \mathbf{e}_{l} +t_2 \cos \mathbf{k}\cdot \mathbf{e}'_{l}), 
\end{align}
for the square lattice [$\mathbf{e}_1=\hat{\mathbf{x}}=(1,0)$, $\mathbf{e}_2=\hat{\mathbf{y}}=(0,1)$, $\mathbf{e}'_1=\hat{\mathbf{x}}+\hat{\mathbf{y}}$, and $\mathbf{e}'_2=\hat{\mathbf{x}}-\hat{\mathbf{y}}$] and 
\begin{align}
\varepsilon_{\mathbf{k}} = -2\sum_{l=1,2, 3}(t_1 \cos \mathbf{k}\cdot \mathbf{e}_{l} +t_2 \cos \mathbf{k}\cdot \mathbf{e}'_{l}), 
\end{align}
for the triangular lattice ($\mathbf{e}_1=\hat{\mathbf{x}}$, $\mathbf{e}_2=-\hat{\mathbf{x}}/2+\sqrt{3}\hat{\mathbf{y}}/2$, $\mathbf{e}_3=-\hat{\mathbf{x}}/2-\sqrt{3}\hat{\mathbf{y}}/2$,
$\mathbf{e}'_1=\mathbf{e}_2+2\mathbf{e}_3$, $\mathbf{e}'_2=\mathbf{e}_3+2\mathbf{e}_1$, and $\mathbf{e}'_3=\mathbf{e}_1+2\mathbf{e}_2$). 
We set the lattice constant $a=1$ as the length unit.

\begin{figure}[t!]
\begin{center}
\includegraphics[width=1.0 \hsize]{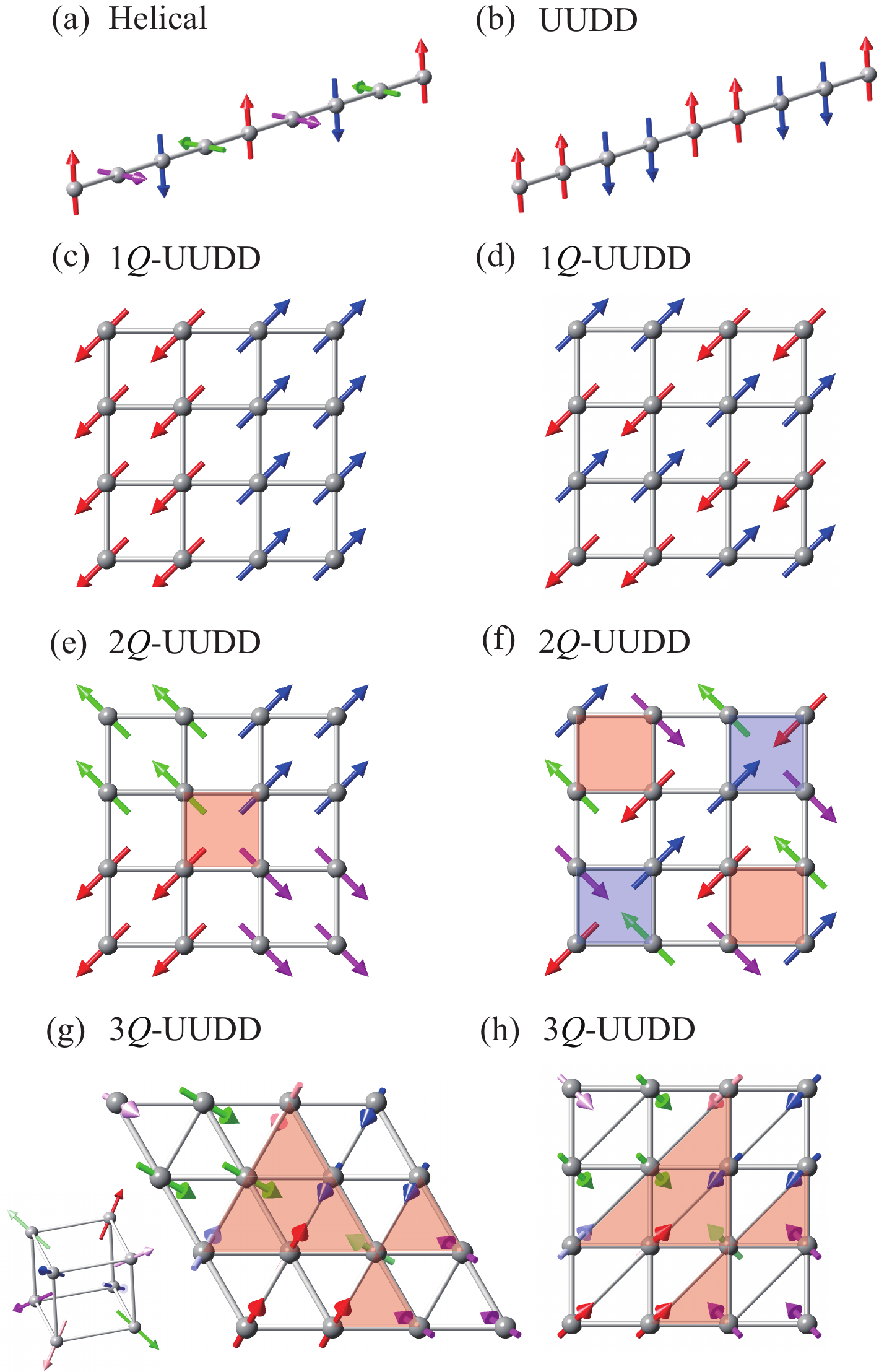} 
\caption{
\label{Fig:multipleQ_ponti}
(Color online)
Schematic pictures of (a) helical and (b) UUDD magnetic structures in the one-dimensional representation, (c) collinear single-$Q$ (1$Q$) UUDD, (d) 1$Q$ UUDD with different ordering vectors from (c) (see the text in detail), (e) coplanar double-$Q$ (2$Q$) UUDD consisting of a superposition of (c), (f) 2$Q$ UUDD consisting of a superposition of (d) on the square lattice, and (g) noncoplanar triple-$Q$ (3$Q$) UUDD magnetic structures on the triangular lattice. 
The arrows denote the directions of localized moments. 
The inset of (g) shows the directions of magnetic moments in the 3$Q$ UUDD state; each spin points along the local [111] directions in the cubic representation. 
In all cases, a global spin rotation does not alter the consequences due to the SU(2) symmetry in the system. 
(h) The square lattice with diagonal bonds, which is topologically equivalent to the triangular lattice in (g). 
In (e) and (f) [(g) and (h)], the red and blue plaquettes show the positive and negative vector (scalar) chirality defined in Eq.~(\ref{eq:vecchi}) [Eq.~(\ref{eq:scachi})]. 
}
\end{center}
\end{figure}

\subsection{RKKY Interaction}
\label{sec:RKKY Interaction}

In general, the Kondo lattice model in Eq.~(\ref{eq:Ham_krep}) exhibits magnetic ordering in the ground state. 
The stable spin pattern depends on the electron filling $n=(1/N)\sum_{i\sigma}\langle c_{i\sigma}^\dagger c_{i\sigma}\rangle$ as well as the exchange coupling constant $J$ ($N$ is the number of lattice sites). 
When $J$ is much larger than $t_1$ and $t_2$, the system shows a ferromagnetic order for general electron filling, by the double-exchange ferromagnetic interaction between localized spins induced by the kinetic motion of itinerant electrons~\cite{Zener_PhysRev.82.403,*anderson1955considerations}. 
On the other hand, when $J \ll t_1$ and $t_2$, the magnetic ordering in the ground state is predominantly determined by the RKKY interaction, which is also a kinetically-induced effective magnetic interaction~\cite{Ruderman,*Kasuya,*Yosida1957}. 
The expression of the RKKY interaction is obtained by the second-order perturbative expansion with respect to $J$ as 
\begin{align}
\label{eq:RKKYHam}
\mathcal{H}^{(2)}=-J^2 \sum_{\mathbf{q}}\chi_{\mathbf{q}}^0 |S_{\mathbf{q}}|^2, 
\end{align}
where $\chi_{\mathbf{q}}^0$ is the bare susceptibility of itinerant electrons, 
\begin{align}
\label{eq:chi0}
\chi_{\mathbf{q}}^0 = \frac{1}{N} \sum_{\mathbf{k}} \frac{f(\varepsilon_{\mathbf{k}})-f(\varepsilon_{\mathbf{k}+\mathbf{q}})}{\varepsilon_{\mathbf{k}+\mathbf{q}}-\varepsilon_{\mathbf{k}}}. 
\end{align}
Here, $f(\varepsilon_{\mathbf{k}})$ is the Fermi distribution function. 
As the bare susceptibility depends on the transfer integrals (noninteracting band structure) and chemical potential (electron filling), these two factors play a decisive role in determining the magnetic state in the Kondo lattice model for $J \ll t_1, t_2$.  

In general, the RKKY interaction in Eq.~(\ref{eq:RKKYHam}) favors a coplanar helical magnetic order, 
whose spin pattern is given by
\begin{align}
\label{eq:1Q_helical}
\mathbf{S}_i=(\cos \mathbf{Q}_1 \cdot \mathbf{r}_i, 0, \sin \mathbf{Q}_1 \cdot \mathbf{r}_i). 
\end{align}
Note that the helical plane, which is taken as the $xz$ plane, is arbitrary in the model with isotropic exchange interactions. 
The ordering vector $\mathbf{Q}_1$ is determined by the maximum of $\chi^0_{\mathbf{q}}$, and therefore, depends on the band structure and electron filling. 
The preference of the helical spin state is understood from the constraint $|\mathbf{S}_i|=1$ and the sum rule $\sum_{\mathbf{q}}|\mathbf{S}_{\mathbf{q}}|^2=N$. 
Other complicated magnetic structures, such as the superpositions of the helical spin patterns, need higher harmonics in order to satisfy the constraint of $|\mathbf{S}_i|=1$, which leads to an energy cost under the sum rule $\sum_{\mathbf{q}}|\mathbf{S}_{\mathbf{q}}|^2=N$. 

\subsection{UUDD Magnetic Structure}
\label{sec:UUDD Magnetic Structure}

For particular ordering vectors, however, magnetic patterns, which are more complicated than the helical one, are allowed without introducing higher harmonics. 
An example is the multiple-$Q$ state which is composed of a superposition of different 1$Q$ states. 
For instance, the double-$Q$ (2$Q$) state with $\mathbf{Q}_1=(\pi,0)$ and $\mathbf{Q}_2=(0,\pi)$ on the square lattice, whose spin structure is given by~\cite{Agterberg_PhysRevB.62.13816,Hayami_PhysRevB.90.060402} 
\begin{align}
\label{eq:spinflux}
\mathbf{S}_i= \hat{\mathbf{x}}\cos \mathbf{Q}_1 \cdot \mathbf{r}_i+\hat{\mathbf{y}}\cos \mathbf{Q}_2 \cdot \mathbf{r}_i,  
\end{align}
satisfies the constraint $|\mathbf{S}_i|=1$. 
The important point is that the modulation with the second component $\mathbf{Q}_2$ is introduced in the perpendicular direction to that of $\mathbf{Q}_1$; this guarantees no additional energy cost at the RKKY level in Eq.~(\ref{eq:RKKYHam}). 
Thus, the 1$Q$ helical state in Eq.~(\ref{eq:1Q_helical}) and the 2$Q$ state in Eq.~(\ref{eq:spinflux}) are energetically degenerate within the RKKY level. 
This indicates that the RKKY interaction is not sufficient to determine the ground state, when $\mathbf{Q}_1=(\pi,0)$ and $\mathbf{Q}_2=(0,\pi)$ maximize the bare susceptibility. 
Note that $\mathbf{Q}_1$ and $\mathbf{Q}_2$ are related with each other by the fourfold rotational symmetry, which is compatible with the square lattice. 
The stability of this type of multiple-$Q$ states was discussed in Refs.~\onlinecite{Hayami_PhysRevB.90.060402,hayami_PhysRevB.91.075104}. 

A similar but different situation can occur for $\mathbf{Q}_1 = (\pi/2,0)$. 
Interestingly, for this particular wave number, there are energetically-degenerate states even within the 1$Q$ states: 
The helical order in Eq.~(\ref{eq:1Q_helical}) with $\mathbf{Q}_1=(\pi/2,0)$ [Fig.~\ref{Fig:multipleQ_ponti}(a)] has the same energy as a collinear UUDD order, whose spin texture is represented by 
\begin{align}
\label{eq:spinuudd}
\mathbf{S}_i = \left[\cos \mathbf{Q}_1 \cdot \mathbf{r}_i+ \cos (\mathbf{Q}_1 \cdot \mathbf{r}_i- \pi/2) \right] \hat{\mathbf{x}}. 
\end{align}
One can easily find that Eq.~(\ref{eq:spinuudd}) satisfies $|\mathbf{S}_i|=1$ as the helical spin structure does. 
The one- and two-dimensional examples are shown in Figs.~\ref{Fig:multipleQ_ponti}(b) and \ref{Fig:multipleQ_ponti}(c), respectively. 
In the two-dimensional case, we can also find another UUDD state with $\mathbf{Q}_1=(\pi/2,\pi)$, as shown in Fig.~\ref{Fig:multipleQ_ponti}(d). 
These UUDD states can be regarded as the superpositions of the helical states, i.e., the spin structures in the UUDD states are decomposed into a pair of $\exp(i \mathbf{Q}_1 \cdot \mathbf{r})$ and $\exp(-i \mathbf{Q}_1 \cdot \mathbf{r})$. 
In the following, we will discuss the stability and nature of 1$Q$ UUDD and multiple-$Q$ UUDD states, which are introduced in the next section, in comparison with the helical state. 

\subsection{Multiple-$Q$ UUDD States} 
\label{sec:Spin and Chirality Configurations for Multiple-$Q$ UUDD States}

We can extend the UUDD states by considering the multiple-$Q$ superpositions. 
In a similar way to Eq.~(\ref{eq:spinflux}), we can define the 2$Q$ UUDD state as 
\begin{align}
\label{eq:spin2QUUDD}
\mathbf{S}_i  = 
 \frac{1}{\sqrt{2}} \left(
\begin{array}{c} 
\cos \mathbf{Q}_1 \cdot \mathbf{r}_i+ \cos (\mathbf{Q}_1 \cdot \mathbf{r}_i- \pi/2) \\ 
\cos \mathbf{Q}_2 \cdot \mathbf{r}_i+ \cos (\mathbf{Q}_2 \cdot \mathbf{r}_i- \pi/2 ) \\ 
0 \\ 
\end{array}
\right). 
\end{align}
In the case of the square lattice, there are two combinations of ordering vectors, which are allowed for the 2$Q$ UUDD state while keeping $|\mathbf{S}_i|=1$: 
One is $\mathbf{Q}_1=(\pi/2,0)$ and $\mathbf{Q}_2=(0,\pi/2)$, and the other is $\mathbf{Q}_1=(\pi/2,\pi)$ and $\mathbf{Q}_2=(\pi,-\pi/2)$. 
Note that, in each combination, $\mathbf{Q}_1$ and $\mathbf{Q}_2$ are connected by the fourfold rotational operation compatible with the square lattice. 
When we choose $\mathbf{Q}_1=(\pi/2,0)$ and $\mathbf{Q}_2=(0,\pi/2)$, the real-space spin configuration is schematically shown in Fig.~\ref{Fig:multipleQ_ponti}(e), while that for $\mathbf{Q}_1=(\pi/2,\pi)$ and $\mathbf{Q}_2=(\pi,-\pi/2)$ is shown in Fig.~\ref{Fig:multipleQ_ponti}(f). 
Their spin configurations are noncollinear but coplanar. 

Meanwhile, we can also consider the triple-$Q$ (3$Q$) UUDD state on the triangular lattice, whose spin configuration is given by 
\begin{align}
\label{eq:spin3QUUDD}
\mathbf{S}_i  = 
 \frac{1}{\sqrt{3}} \left(
\begin{array}{l} 
\cos \mathbf{Q}_1 \cdot \mathbf{r}_i+ \cos (\mathbf{Q}_1 \cdot \mathbf{r}_i-\pi/2) \\ 
\cos \mathbf{Q}_2 \cdot \mathbf{r}_i+ \cos (\mathbf{Q}_2 \cdot \mathbf{r}_i-\pi/2 ) \\ 
 \cos \mathbf{Q}_3 \cdot \mathbf{r}_i+ \cos ( \mathbf{Q}_3 \cdot \mathbf{r}_i-\pi/2)  \\ 
\end{array}
\right),
\end{align}
where $\mathbf{Q}_1=(\pi/2,0)$, $\mathbf{Q}_2=(0,-\pi/2)$, and $\mathbf{Q}_3=(-\pi/2,\pi/2)$.  
Here and hereafter, we regard the triangular lattice as a topologically equivalent square lattice with diagonal bonds, as shown in Fig.~\ref{Fig:multipleQ_ponti}(h).  
The spin configuration given by Eq.~(\ref{eq:spin3QUUDD}) is noncoplanar, whose original geometry is shown in Fig.~\ref{Fig:multipleQ_ponti}(g) ($\mathbf{Q}_1$, $\mathbf{Q}_2$, and $\mathbf{Q}_3$ are connected by the sixfold rotational operation compatible with the triangular lattice). 

As exemplified in Sec.~\ref{sec:UUDD Magnetic Structure} for the case of $\mathbf{Q}_1=(\pi,0)$ and $\mathbf{Q}_2 = (0,\pi)$, the RKKY energy for a multiple-$Q$ state remains the same as that in the 1$Q$ state when there are no higher harmonics in the spin configurations. 
Hence, the multiple-$Q$ UUDD states introduced in this section have the same RKKY energy as those for the 1$Q$ helical and UUDD states. 
The degeneracy is lifted by higher-order contributions beyond the RKKY interaction, as discussed in Sec.~\ref{sec:Instability toward Multiple-$Q$ UUDD}. 

\subsection{Chirality Density Waves}
\label{sec:Chirality Density Waves}

The multiple-$Q$ UUDD states exhibit nonzero chirality. 
We define the vector and scalar chirality as 
\begin{align}
\label{eq:vecchi}
\bm{\chi}_{\rm v}^p&= \frac{1}{4}(\mathbf{S}_{i}\times \mathbf{S}_j+\mathbf{S}_{j}\times \mathbf{S}_k+\mathbf{S}_{k}\times \mathbf{S}_l+\mathbf{S}_{l}\times \mathbf{S}_i), \\
\label{eq:scachi}
\chi_{\rm s}^p&= \mathbf{S}_m \cdot (\mathbf{S}_{n}\times \mathbf{S}_o), 
\end{align}
respectively, where $i$, $j$, $k$, and $l$ ($m$, $n$, and $o$) are sites on each square (triangle) plaquette $p$ in a counterclockwise direction. 
In the multiple-$Q$ UUDD states, the value of chirality depends on the spatial position of the plaquette, which we call ChDW. 

Indeed, in the 2$Q$ UUDD on the square lattice [Eq.~(\ref{eq:spin2QUUDD})], the $z$ component of the vector chirality $\bm{\chi}_{\rm v}^p$ oscillates from a positive to negative value on each plaquette, as shown in Figs.~\ref{Fig:multipleQ_ponti}(e) and \ref{Fig:multipleQ_ponti}(f); see also in Figs.~\ref{Fig:KPM}(a) and \ref{Fig:edge}(a) for the chirality distribution in a larger scale. 
Thus, this state is an antiferroic-type vector ChDW with vanishing net vector chirality. 
Note that the scalar chirality is zero everywhere because of the coplanar spin configurations. 
Meanwhile, in the 3$Q$-UUDD state, the scalar chirality takes a positive value or zero in a periodic way, as shown in Figs.~\ref{Fig:multipleQ_ponti}(g) and \ref{Fig:multipleQ_ponti}(h). 
This is a ferri-type scalar ChDW with a nonzero net scalar chirality. 
(In this noncoplanar 3$Q$ case, we do not discuss the vector chirality.) 

Thus, these ChDW states are distinct from the ferroic chirality orders in the previous studies, where every plaquette possesses the same value of vector or scalar chirality, as mentioned in the introduction~\cite{Martin_PhysRevLett.101.156402,Shindou_PhysRevLett.87.116801}. 
Furthermore, they have richer superstructures in the chirality than the one-dimensional stripy ones in the previous studies~\cite{Solenov_PhysRevLett.108.096403,ozawa2015vortex}. 
Reflecting the distinct aspect, intriguing edge-dependent electronic structures are obtained as discussed in Sec.~\ref{sec:Electronic Structure}. 

\section{Instability toward Multiple-$Q$ UUDD States}
\label{sec:Instability toward Multiple-$Q$ UUDD}

In this section, we examine the instability toward the multiple-$Q$ UUDD states from the energetic point of view. 
In Sec.~\ref{sec:Perturbative Analysis}, we show the results from higher-order perturbative expansion with respect to $J$ beyond the second-order RKKY contribution. 
In Sec.~\ref{sec:Variational Calculation}, we evaluate the energy differences between the multiple-$Q$ UUDD, 1$Q$ UUDD, and helical states by variational calculations. 
Finally, we examine the ground state in an unbiased way by performing the KPM-LD simulation in Sec.~\ref{sec:Langevin-based Simulation}. 
The results provide complementary evidences that the system has the instability toward the multiple-$Q$ UUDD states at particular electron fillings. 

\subsection{Perturbative Analysis}
\label{sec:Perturbative Analysis}

\begin{figure}[t!]
\begin{center}
\includegraphics[width=1.0 \hsize]{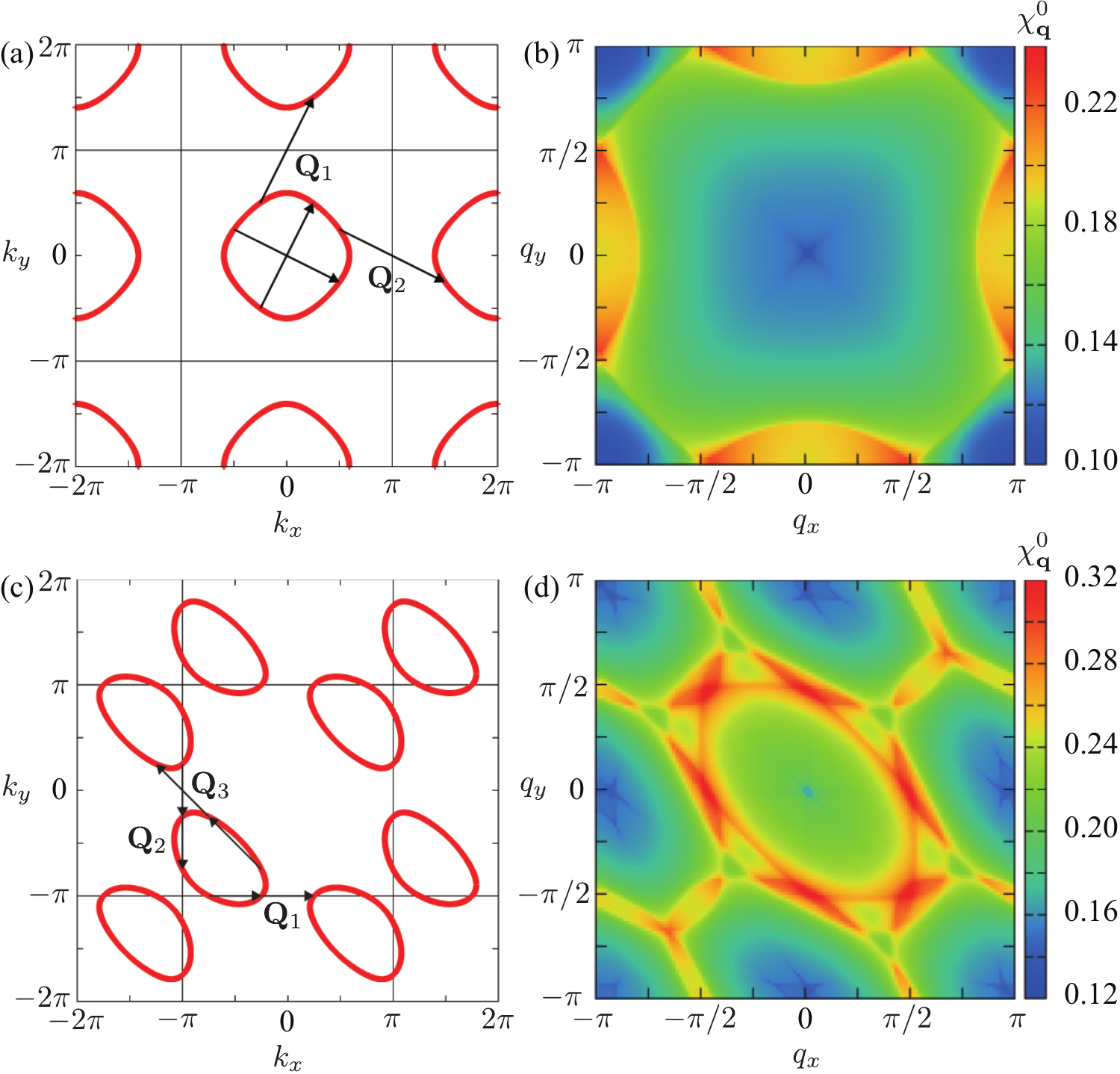} 
\caption{
\label{Fig:suscep}
(Color online)
The Fermi surface for (a) the square lattice model at $t_2=0$ and $\mu=-\sqrt{2}$ and (c) triangular lattice model at 
$t_2=0.5$ and $\mu=2$. 
The triangular lattice is regarded as a topologically equivalent square lattice with diagonal bonds, as shown in Fig.~\ref{Fig:multipleQ_ponti}(h). 
$\mathbf{Q}_\nu$ ($\nu=1$, 2, and 3) are the vectors connecting the Fermi surfaces. 
(b), (d) The contour plots of the bare susceptibility corresponding to (a) and (c), respectively. 
The bare susceptibility exhibits maxima at $\mathbf{Q}_\nu$ and the symmetry-related wave numbers. 
}
\end{center}
\end{figure}

As described in the previous section, when particular symmetry-related wave numbers maximize the bare susceptibility, the RKKY interaction is not sufficient to determine the ground state, since it gives the same energy for helical, 1$Q$ UUDD, and multiple-$Q$ UUDD orders. 
This occurs when the ordering vectors satisfy $|Q_\xi|=\pi$, $\pi/2$, or $0$ [$\mathbf{Q}=(Q_x,Q_y)\neq \mathbf{0}$ or $(\pi,\pi)$], and they are related with each other by the rotational operation proper to the lattice structure. 

An example is found for the square lattice model at $t_2=0$ and the chemical potential $\mu=-\sqrt{2}$. 
The Fermi surface is drawn in Fig.~\ref{Fig:suscep}(a). 
As shown in the figure, the Fermi surface is connected by two wave numbers, $\mathbf{Q}_1=(\pi/2,\pi)$ and $\mathbf{Q}_2=(\pi,-\pi/2)$, which satisfy the above condition. 
Note that the connections are not only within the same Brillouin zone but also between different Brillouin zones. 
These connections maximize the bare susceptibility at $\mathbf{Q}_1$, $\mathbf{Q}_2$, and the symmetry-related wave numbers, $-\mathbf{Q}_1$ and $-\mathbf{Q}_2$, as shown in Fig.~\ref{Fig:suscep}(b). 
This indicates that the magnetically ordered states with these ordering vectors are the candidates for the ground state. 
Specifically, the plausible ground states are the 1$Q$ helical state with $\mathbf{Q}_1$, the 1$Q$ UUDD with $\mathbf{Q}_1$, and the 2$Q$ UUDD with $\mathbf{Q}_1$ and $\mathbf{Q}_2$; these three states are energetically degenerate for the RKKY Hamiltonian in Eq.~(\ref{eq:RKKYHam}). 
Note that we do not need to consider the 3$Q$ UUDD state in this square lattice case, because its RKKY energy is higher than that for the 1$Q$ UUDD state; the 3$Q$ state becomes relevant in the triangular-lattice system with sixfold rotational symmetry. 

In fact, another example including the possibility of 3$Q$ UUDD ordering is found for the triangular lattice model at $t_2=0.5$ and $\mu=2$. 
Figure~\ref{Fig:suscep}(c) shows the Fermi surface. 
In this case, there are three vectors connecting the Fermi surface, $\mathbf{Q}_1=(\pi/2,0)$, $\mathbf{Q}_2=(0,-\pi/2)$, and $\mathbf{Q}_3=(-\pi/2,\pi/2)$, which lead to the six maxima in the bare susceptibility shown in Fig.~\ref{Fig:suscep}(d). 
Thus, the possible ground states in this case include the 3$Q$ UUDD with $\mathbf{Q}_1$, $\mathbf{Q}_2$, and $\mathbf{Q}_3$ in addition to the 1$Q$ and 2$Q$ states. 

Similar connections of the Fermi surface occur for $\mathbf{Q}_1=(\pi/2,0)$ and $\mathbf{Q}_2=(0,\pi/2)$ at $\mu=-\sqrt{2}(1+\sqrt{2})$ on the square lattice model with $t_2=0$, and for $\mathbf{Q}_1=(\pi/2,0)$, $\mathbf{Q}_2=(0,-\pi/2)$, and $\mathbf{Q}_3=(-\pi/2,\pi/2)$ at $\mu=-2(1+\sqrt{2})$ on the triangular lattice model with $t_2=0$. 
In these two cases, although $\chi_\mathbf{q}^0$ has maxima at $\pm\mathbf{Q}_\nu$, it shows less $\mathbf{q}$ dependence and the peaks are not clearly visible (not shown here), compared to the previous cases shown in Figs.~\ref{Fig:suscep}(b) and \ref{Fig:suscep}(d). 
Nevertheless, we will discuss these two cases in addition to the former two, as the perturbative arguments below indicate that the instability toward the multiple-$Q$ states appears in a common manner. 

\begin{figure}[t!]
\begin{center}
\includegraphics[width=1.0 \hsize]{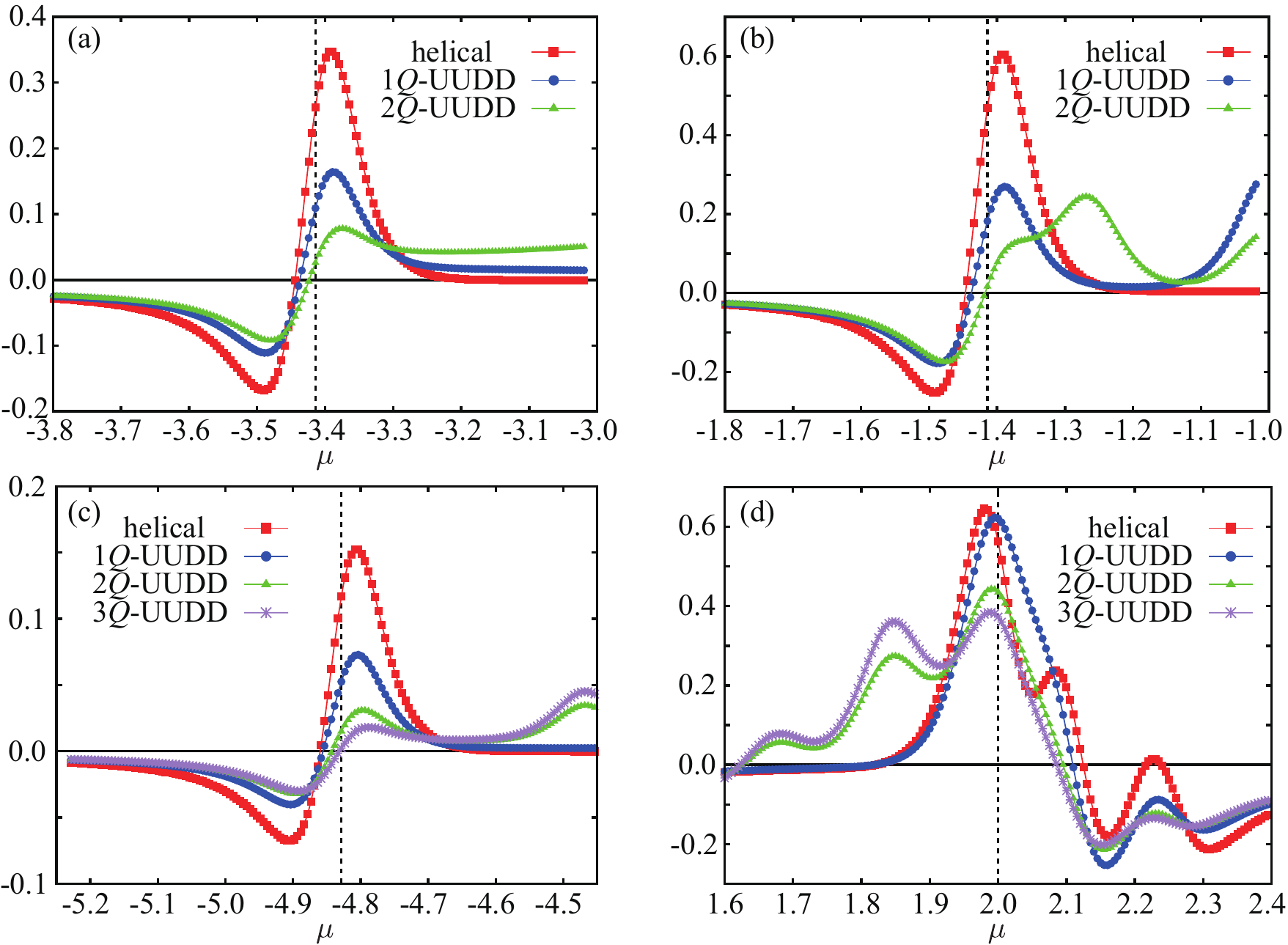} 
\caption{
\label{Fig:perturbation}
(Color online)
The fourth-order contributions to the free energy, $F_{\nu}^{(4)}$ and $F_{\rm helical}^{(4)}$ in Eqs.~(\ref{eq:F4UUDD}) and (\ref{eq:freeenergy}), respectively, divided by $J^4$, as functions of the chemical potential $\mu$ for the ground state candidates at the RKKY level on (a), (b) square, and (c), (d) triangular lattices. 
The parameters are (a) $t_2=0$, $\mathbf{Q}_1=(\pi/2,0)$, and $\mathbf{Q}_2=(0,\pi/2)$, (b) $t_2=0$, $\mathbf{Q}_1=(\pi/2,\pi)$, and $\mathbf{Q}_2=(\pi,-\pi/2)$,  (c) $t_2=0$, $\mathbf{Q}_1=(\pi/2,0)$, $\mathbf{Q}_2=(0,-\pi/2)$, and $\mathbf{Q}_3=(-\pi/2,\pi/2)$, and  (d) $t_2=0.5$, $\mathbf{Q}_1=(\pi/2,0)$, $\mathbf{Q}_2=(0,-\pi/2)$, and $\mathbf{Q}_3=(-\pi/2,\pi/2)$. 
The data are calculated at $T=0.03$. 
The vertical dashed lines point to the optimal chemical potential where the bare susceptibility has maxima at the corresponding wave numbers: (a) $\mu = -\sqrt{2}(1+\sqrt{2})$, (b) $-\sqrt{2}$, (c) $-2(1+\sqrt{2})$, and (d) $2$. 
}
\end{center}
\end{figure}

For the situations above, the RKKY energy is degenerate for the ground state candidates. 
Hence, in order to discriminate them, we need to take into account the higher-order contributions in the perturbative analysis~\cite{Akagi_PhysRevLett.108.096401,Hayami_PhysRevB.90.060402}. 
The degeneracy at the second-order RKKY level is lifted by the fourth-order contribution with respect to $J$ (note that the odd-order terms vanish by symmetry). 
When we expand the free energy $F=-T \sum_{\mu} {\rm log} \left[ 1+ \exp (-E_\mu/T)\right]$ ($E_\mu$ are the eigenvalues of $\mathcal{H} $ and $T$ is temperature) as $F=F^{(0)}+F^{(2)}+F^{(4)} \cdots$, the fourth-order contribution $F^{(4)}$ for the four possible candidates are analytically obtained as 
\begin{align}
\label{eq:F4UUDD}
F^{(4)}_{\nu} &=\frac{J^4}{2}\left\{ \left(1-\frac{1}{\nu}\right) A + \frac{1}{2\nu} B \right\}, \\
\label{eq:freeenergy}
F^{(4)}_{\rm helical} &=\frac{J^4}{2}T\sum_{\mathbf{k} \omega_p}  G_{\mathbf{k}+\mathbf{Q}}^2 G_{\mathbf{k}}^2,  
\end{align}
where 
\begin{align}
A&=T\sum_{\mathbf{k} \omega_p} (G_{\mathbf{k}}^2 G_{\mathbf{k}+\mathbf{Q}_{\eta}}G_{\mathbf{k}+\mathbf{Q}_{\eta'}}+G_{\mathbf{k}} G^2_{\mathbf{k}+\mathbf{Q}_{\eta}}G_{\mathbf{k}+\mathbf{Q}_{\eta}+\mathbf{Q}_{\eta'}} \nonumber \\
&-G_{\mathbf{k}} G_{\mathbf{k}+\mathbf{Q}_{\eta}}G_{\mathbf{k}+\mathbf{Q}_{\eta'}}G_{\mathbf{k}+\mathbf{Q}_{\eta}+\mathbf{Q}_{\eta'}}), 
\label{eq:A}
\\
B&=T\sum_{\mathbf{k} \omega_p} 
(G_{\mathbf{k}}^2 G_{\mathbf{k}+\mathbf{Q}_{\eta}}^2-G_{\mathbf{k}} G_{\mathbf{k}+\mathbf{Q}_{\eta}}G_{\mathbf{k}+2\mathbf{Q}_{\eta}}G_{\mathbf{k}+3\mathbf{Q}_{\eta}} \nonumber \\
&+2 G_{\mathbf{k}} G^2_{\mathbf{k}+\mathbf{Q}_{\eta}}G_{\mathbf{k}+2\mathbf{Q}_{\eta}}
).
\label{eq:B}
\end{align}
In Eq.~(\ref{eq:F4UUDD}), $\nu=1$, 2, and 3 stand for the 1$Q$, 2$Q$, and 3$Q$ UUDD states, and $G_{\mathbf{k}} (i \omega_p) = \left[ i \omega_p -(\varepsilon_{\mathbf{k}}-\mu) \right]^{-1}$ is noninteracting Green's function, where $\omega_p$ is the Matsubara frequency. 
In Eqs.~(\ref{eq:A}) and (\ref{eq:B}), $(\eta,\eta')=(1,2)$ for 2$Q$ and $(\eta,\eta')=(1,2)$, $(1,3)$, and $(2,3)$ for 3$Q$.

Figure~\ref{Fig:perturbation} shows $F^{(4)}$ in 
Eqs.~(\ref{eq:F4UUDD}) and (\ref{eq:freeenergy}) while changing the chemical potential $\mu$ around the values for which the RKKY interaction favors magnetic orders with $\mathbf{Q}_{1}=(\pi/2,0)$ or $\mathbf{Q}_{1}=(\pi/2,\pi)$ (the vertical dashed lines in each figure). 
For the square lattice case, we find $ F^{(4)}_{2} <F^{(4)}_{1} <F^{(4)}_{{\rm helical}}$, as shown in Figs.~\ref{Fig:perturbation}(a) and \ref{Fig:perturbation}(b). 
The results indicate that the fourth-order contribution favors the 2$Q$ UUDD states near the particular fillings, where the Fermi surfaces are connected by $\mathbf{Q}_\nu$. 
Specifically, the 2$Q$ UUDD states are favored at the electron filling $n\sim0.097$ [$\mu\sim-\sqrt{2}(1+\sqrt{2})$] in Fig.~\ref{Fig:perturbation}(a) and $n\sim 0.506$ ($\mu\sim\sqrt{2}$) in Fig.~\ref{Fig:perturbation}(b). 
On the other hand, in the triangular-lattice case, we obtain $ F^{(4)}_{3} <F^{(4)}_{2} <F^{(4)}_{1} <F^{(4)}_{{\rm helical}}$, as shown in Fig.~\ref{Fig:perturbation}(c) for $t_2=0$ and Fig.~\ref{Fig:perturbation}(d) for $t_2=0.5$. 
Hence, in these cases, the fourth-order contribution prefers to the 3$Q$ UUDD states: at $n \sim 0.113$ [$\mu\sim-2(1+\sqrt{2})$] in Fig.~\ref{Fig:perturbation}(c) and $n\sim 1.618$ ($\mu\sim 2$) in Fig.~\ref{Fig:perturbation}(d).

Thus, in all cases, higher multiple-$Q$ states are favored by the fourth-order perturbation beyond the RKKY interaction. 
The instability toward the multiple-$Q$ states is understood by the (local) gap formation in the band structure of itinerant electrons due to (partial) nesting of the Fermi surfaces. 
In general, a magnetic order by the Fermi surface nesting opens a gap at the Fermi surfaces connected by the ordering vector. 
The multiple-$Q$ orders have more connections than the 1$Q$ orders, as exemplified in Figs.~\ref{Fig:suscep}(a) and \ref{Fig:suscep}(c). 
The connections therefore lead to a larger energy gain owing to the gap opening at the multiple points on the Fermi surfaces. 
A similar mechanism was discussed for other noncoplanar multiple-$Q$ orders~\cite{Akagi_PhysRevLett.108.096401,Hayami_PhysRevB.90.060402,ozawa2015vortex}. 
Therefore, in the small $J$ limit, a 2$Q$ (3$Q$) UUDD state is expected to be realized on the square (triangular) lattice. 
However, it is worth noting that, although the perturbative arguments above indicate the instabilities toward the multiple-$Q$ orders, the fourth-order corrections in Eqs.~(\ref{eq:F4UUDD}) and (\ref{eq:freeenergy}) diverge in the limit of $T \to 0$. 
This suggests the breakdown of the perturbative theory, and hence, we need to carefully check the validity by complementary methods, such as numerical simulations, as we will discuss in the following sections. 

Let us remark on the comparison between the 1$Q$ UUDD and helical states. 
The fourth-order perturbative analysis shows that the energy for the UUDD state is always lower than the helical one near the particular electron fillings, as shown in Fig.~\ref{Fig:perturbation}. 
This is because the UUDD state has more perturbative processes than the helical one, as shown in Eqs.~(eq:F4UUDD) and (eq:freeenergy). 
The difference is related with the inversion symmetry breaking by the helical order. 
The UUDD order opens a local gap in the band structure owing to the multiple processes, whereas the helical one does not.
The preference of the 1$Q$ UUDD state with $Q=\pi/2$ than the helical one was indeed found in the study for the one-dimensional Kondo lattice model~\cite{minami2015low,matsueda2015comment}.  
Our perturbative arguments not only support the preference but also show that the tendency is general irrespective of the system dimensions.

\subsection{Variational Calculation}
\label{sec:Variational Calculation}

\begin{figure}[t!]
\begin{center}
\includegraphics[width=1.0 \hsize]{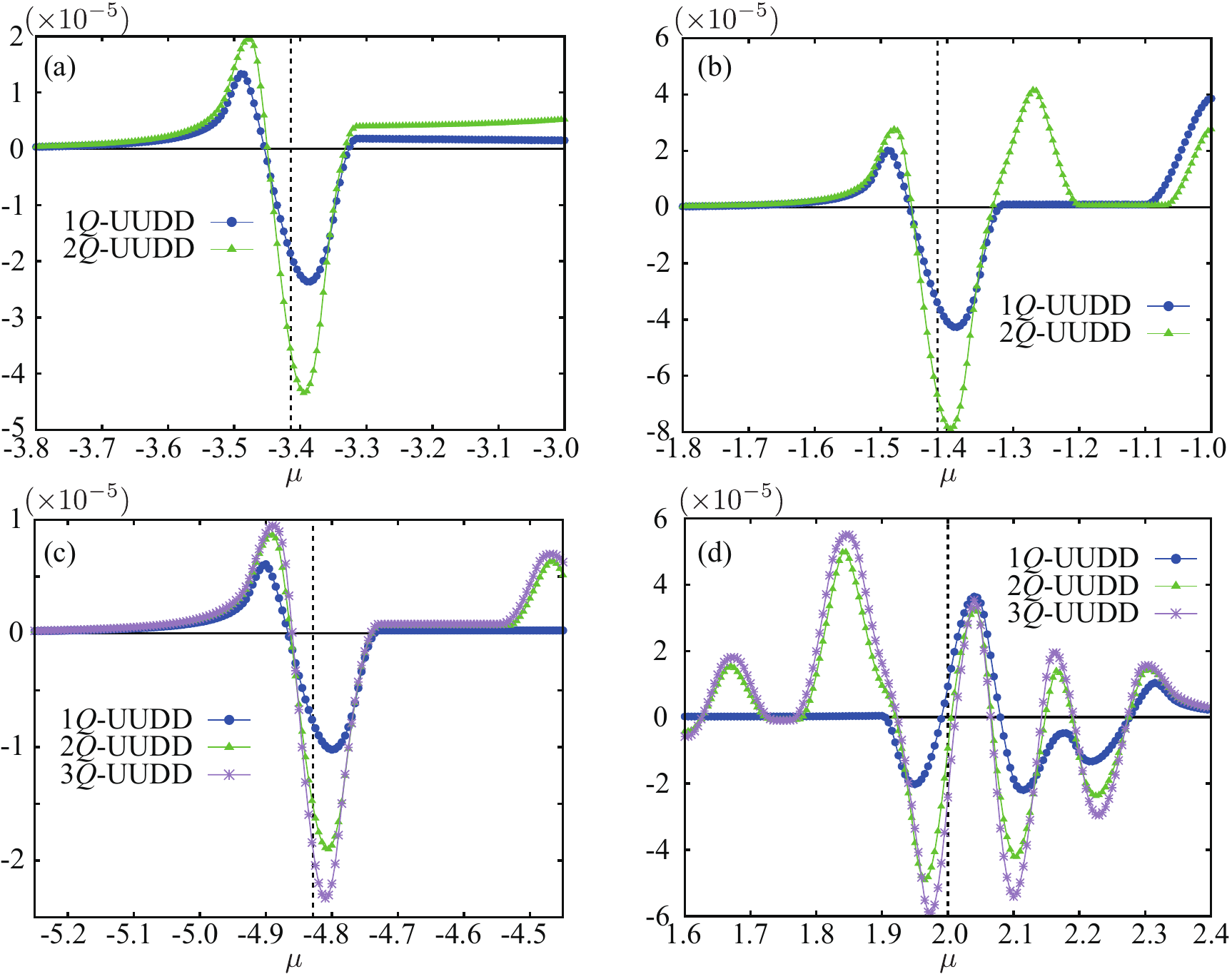} 
\caption{
\label{Fig:variational}
(Color online)
The grand potential at zero temperature as functions of the chemical potential $\mu$ calculated by the variational calculation for the Hamiltonian in Eq.~(\ref{eq:Ham_krep}) at $J=0.1$ on (a), (b) square, and (c), (d) triangular lattices. 
The grand potential for the 1$Q$, 2$Q$, and 3$Q$ UUDD magnetically ordered states is measured from that for the helical state. 
The model parameters in each panel correspond to those in Fig.~\ref{Fig:perturbation}. 
}
\end{center}
\end{figure}

In order to confirm the perturbative analysis, we numerically examine the ground state of the model in Eq.~(\ref{eq:Ham}). 
In this section, we perform a variational calculation: 
We compare the grand potential at zero temperature, $\Omega= E -\mu n$ ($E=\langle \mathcal{H} \rangle/N $ is the internal energy per site), for variational states with different magnetic orders in the localized spins, and determine the lowest energy state. 
For the variational states, we assume the helical, 1$Q$, 2$Q$, and 3$Q$ UUDD states. 
We consider these ordered states in a sixteen-site unit cell ($4\times 4$), and compute $\Omega$ in the system with 1024 supercells under the periodic boundary conditions.

Figure~\ref{Fig:variational} shows $\mu$ dependences of the grand potential at $J=0.1$ for different variational states measured from that for the helical ordering. 
The results support the perturbative results in Fig.~\ref{Fig:perturbation}: 
In Figs.~\ref{Fig:variational}(a)-\ref{Fig:variational}(d), the grand potential for the 2$Q$ (3$Q$) UUDD state gives the lowest energy for the square (triangular) lattice model, in the $\mu$ regions where the fourth-order free energy $F^{(4)}$ becomes lowest for the corresponding state, as shown in Figs.~\ref{Fig:perturbation}(a)-\ref{Fig:perturbation}(d). 
We note that the 3$Q$ UUDD states are also favored at $\mu \sim 2.1$ and $2.25$, but they may be taken over by other states with slightly different ordering vectors determined by the Fermi surface at these values of the chemical potential. 
Thus, the multiple-$Q$ UUDD states are variationally stable near the electron fillings where the fourth-order perturbation signals their instabilities.

\subsection{Langevin Dynamics Simulation}
\label{sec:Langevin-based Simulation}

\begin{figure}[t!]
\begin{center}
\includegraphics[width=1.0 \hsize]{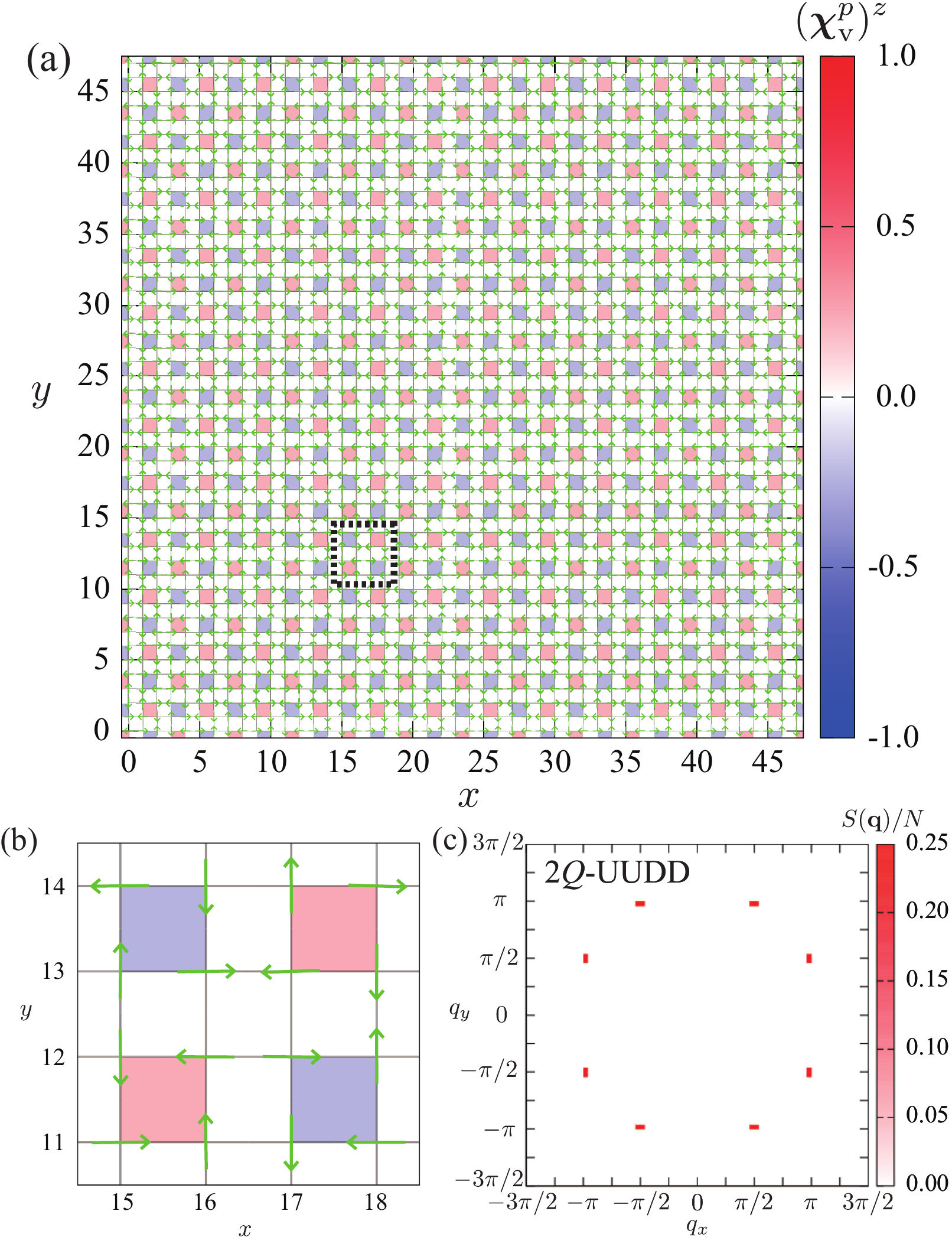} 
\caption{
\label{Fig:KPM}
(Color online)
(a) Real-space configurations of localized spins and the $z$ components of vector chirality, $(\bm{\chi}_{{\rm v}}^p)^z$ [see Eq.~(\ref{eq:vecchi})], in the optimized ground state obtained by the KPM-LD simulation for the model in Eq.~(\ref{eq:Ham}) on the square lattice. 
The simulation is done for a $48^2$-site cluster at $\mu=-1.39$ ($n\simeq0.522$) for $t_2=0$ and $J=0.3$. 
Green arrows at the vertices of the square lattice represent the directions of localized spins, and the color for each square plaquette indicates the value of $(\bm{\chi}_{{\rm v}}^p)^z$. 
(b) Enlarged picture of (a) in the dotted square. 
(c) Spin structure factor divided by the system size obtained from the spin configuration in (a).
}
\end{center}
\end{figure}

For further confirmation of the multiple-$Q$ states, we perform the KPM-LD simulation. 
This is an unbiased numerical simulation based on Langevin dynamics~\cite{Barros_PhysRevB.88.235101}~, in which the kernel polynomial method~\cite{Weis_RevModPhys.78.275} is utilized for enabling the calculations for larger system sizes than the standard Monte Carlo simulation combined with the direct diagonalization.  
We here apply the method to the square lattice model at $\mu\sim-\sqrt{2}$, where the perturbative and variational calculations coherently point to the 2$Q$ UUDD state, as shown in Figs.~\ref{Fig:perturbation}(b) and \ref{Fig:variational}(b), respectively. 
The simulation is done at zero temperature for a $48^2$-site cluster of the square lattice with periodic boundary conditions. 
In the kernel polynomial method, we expand the density of states by up to 4000th order of Chebyshev polynomials with 144 random vectors which are selected by a probing technique~\cite{Tang12}. 
In the Langevin dynamics, we use a projected Heun scheme\cite{Mentink10} for 1000 steps with the time interval  $\Delta \tau =10$.

The results are shown in Fig.~\ref{Fig:KPM}. 
In the simulation, we take a slightly large value of $J$ ($J=0.3$) for ensuring the convergence of numerical optimization. 
Figures~\ref{Fig:KPM}(a) and \ref{Fig:KPM}(b) show a snapshot of the configurations of localized spins and vector chirality [Eq.~(\ref{eq:vecchi})] in the optimized states. 
The obtained state coincides well with the 2$Q$ UUDD order with vector ChDW in Fig.~\ref{Fig:multipleQ_ponti}(f). 
Indeed, it shows eight (two independent) peaks in the spin structure factor, $S(\mathbf{q})= (1/N) \sum_{i,j} \mathbf{S}_i \cdot \mathbf{S}_j e^{i\mathbf{q}\cdot(\mathbf{r}_i-\mathbf{r}_j)}$, as shown in Fig.~\ref{Fig:KPM}(c). 
Thus, the unbiased numerical simulations also support the emergence of multiple-$Q$ UUDD states in the ground state.

\section{Electronic Structure}
\label{sec:Electronic Structure}
In this section, we present the electronic band structures of the multiple-$Q$ UUDD states with ChDW. 
In Sec.~\ref{sec:Bulk Dispersion}, we show that ChDW may bring topological nature in the band structure; e.g., the massless Dirac semimetal and Chern insulator. 
We also show that the edge states in these ChDW states appear in a peculiar way depending on how the ChDW is terminated at the edges in Sec.~\ref{sec:Edge Dispersion}. 
To elucidate the nontrivial electronic structures by the multiple-$Q$ states, hereafter, we assume that the spin configurations in Eqs.~(\ref{eq:spin2QUUDD}) and (\ref{eq:spin3QUUDD}) remain stable for larger $J$. 

\subsection{Bulk Dispersion}
\label{sec:Bulk Dispersion}

\begin{figure}[hbt!]
\begin{center}
\includegraphics[width=1.0 \hsize]{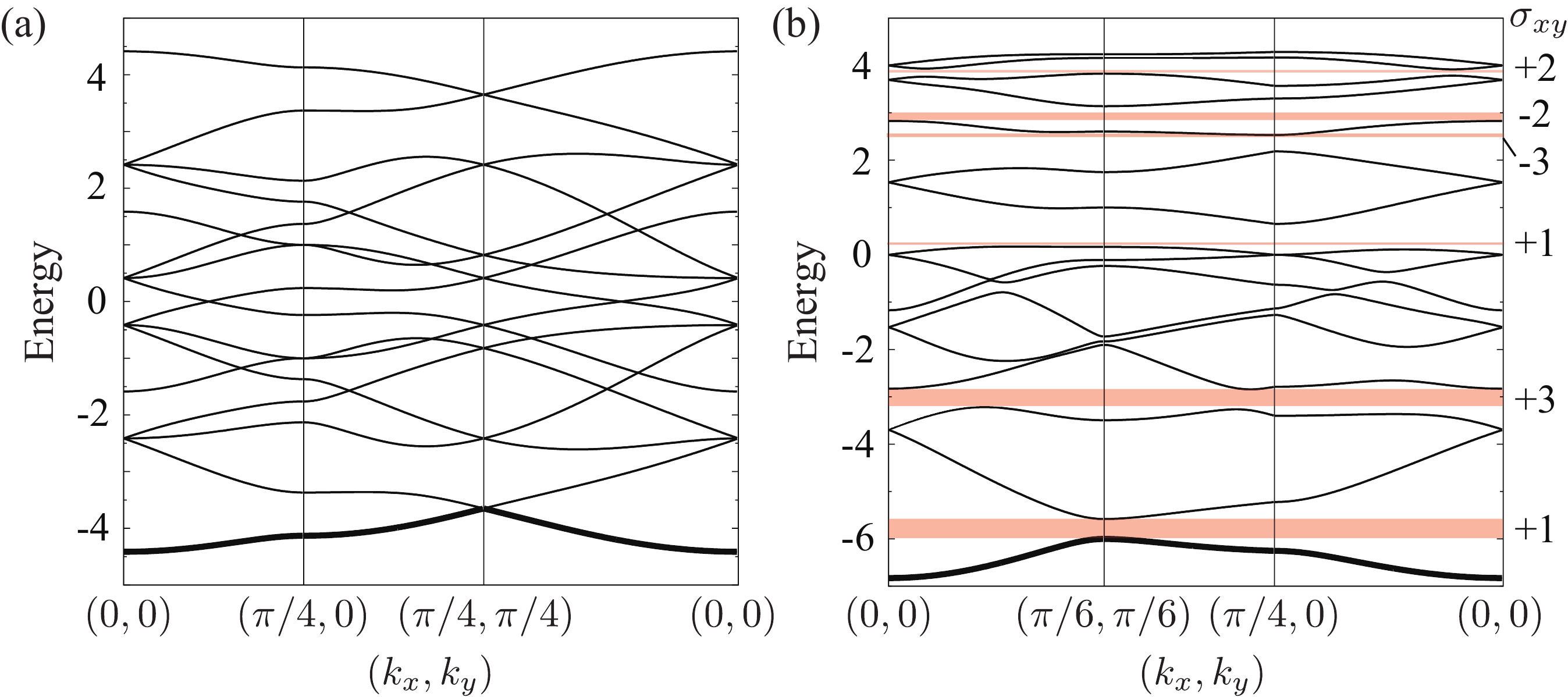} 
\caption{
\label{Fig:band}
(Color online)
Typical energy dispersions of (a) the 2$Q$ UUDD state on the square lattice and (b) the 3$Q$ UUDD state on the triangular lattice. 
The parameters are $t_2=0$, $J=1$, $\mathbf{Q}_1=(\pi/2,0)$, and $\mathbf{Q}_2=(0,\pi/2)$ for the 2$Q$ state, while $t_2=0$, $J=2$, $\mathbf{Q}_1=(\pi/2,0)$, $\mathbf{Q}_2=(0,-\pi/2)$, and $\mathbf{Q}_3=(-\pi/2,\pi/2)$ for the 3$Q$ state. 
The dispersions are shown along the symmetric lines in the folded Brillouin zones. 
The lowest thick curves show the occupied bands at $n=0.125$. 
In (a), the massless Dirac node appears at $\mathbf{k}=(k_x, k_y)=(\pi/4,\pi/4)$ at $n=0.125$ as well as several other fillings. 
Meanwhile, the band gap opens at $n=0.125$ in (b), in addition to several other commensurate fillings. 
In (b), the values of the quantized topological Hall conductivity in unit of $e^2/h$, which are obtained when the Fermi level locates inside the gap, are also shown in the right side of the panel. 
}
\end{center}
\end{figure}

The ChDW associated with the multiple-$Q$ UUDD state may modulate the electronic structure in a nontrivial way through the spin Berry phase mechanism. 
This is demonstrated in Fig.~\ref{Fig:band}. 
Figure~\ref{Fig:band}(a) shows a typical energy dispersion for the 2$Q$ UUDD state with $\mathbf{Q}_1=(\pi/2,0)$, and $\mathbf{Q}_2=(0,\pi/2)$ on the square lattice ($t_2=0$ and $J=1$). 
In this case, the spin texture induces an antiferroic-type ChDW, as shown in Fig.~\ref{Fig:multipleQ_ponti}(e) [see also Fig.~\ref{Fig:edge}(a)]. 
There are sixteen bands and each band is doubly degenerate. 
As shown in the figure, the lowest band plotted by thick curves touches the higher band at the single point at $(\pi/4,\pi/4)$, forming a linear dispersion. 
This is a massless Dirac node, whose electron filling corresponds to $n=0.125$. 
Thus, the $2Q$ UUDD state with the antiferroic ChDW is a Dirac semimetal at $n=0.125$. 
The formation of the Dirac node implies that the 2$Q$ UUDD state may be stabilized at this commensurate filling at nonzero $J$, although the instability in the weak $J$ limit occurs at a slightly smaller filling, $n\sim0.097$, as discussed in the previous sections. 
Similar stabilization with forming Dirac semimetal was discussed for square and cubic lattices~\cite{Agterberg_PhysRevB.62.13816,hayami_PhysRevB.91.075104,Hayami_PhysRevB.89.085124}. 

On the other hand, the other 2$Q$ UUDD state in Fig.~\ref{Fig:multipleQ_ponti}(f) does not show such Dirac nodes; the bands are separated by energy gaps, and the system is a trivial band insulator at $n=0.5$, $1.0$, and $1.5$ (not shown here). 
The difference is understood by considering the $J  \to \infty$ limit: 
in the case of Fig.~\ref{Fig:multipleQ_ponti}(f), itinerant electrons are confined in each four-site plaquette with nonzero vector chirality because of the antiparallel spin configuration between the plaquettes, whereas they are not in Fig.~\ref{Fig:multipleQ_ponti}(e). 

Figure~\ref{Fig:band}(b) represents a typical energy dispersion for the 3$Q$ UUDD state with $\mathbf{Q}_1=(\pi/2,0)$, $\mathbf{Q}_2=(0,-\pi/2)$, and $\mathbf{Q}_3=(-\pi/2,\pi/2)$ on the triangular lattice ($t_2=0$ and $J=2$), which accompanies a partially ferroic-type ChDW, as shown in Figs.~\ref{Fig:multipleQ_ponti}(g) and \ref{Fig:multipleQ_ponti}(h). 
As shown in the figure, the partially ferroic ChDW leads to a gap opening at the Fermi level for $n=0.125$, which is close at $n\sim 0.113$ where the instability toward the 3$Q$ UUDD is anticipated in the small $J$ limit. 
Similar to the square-lattice case above, the gap opening suggests that the 3$Q$ UUDD state may be stable at $n=0.125$ for finite $J$. 
Similar stabilization by gap opening was discussed in Refs.~\onlinecite{Akagi_JPSJ.79.083711,Akagi_PhysRevLett.108.096401}. 
We find that the insulating state is a topologically nontrivial Chern insulator; the lowest band acquires the Chern number $+1$, leading to the quantized topological Hall conductivity, $\sigma_{xy} = e^2/h$ ($e$ is the elementary charge and $h$ is the Planck constant). 
Hence, the 3$Q$ UUDD state with partially ferroic ChDW provides a Chern insulator. 
Note that there are other gaps in states with higher fillings, and the bands separated by the gaps are assigned by the corresponding Chern numbers, as shown in the right side of Fig.~\ref{Fig:band}(b). 
Similar Chern insulators were discussed for ferroic ChDW in noncoplanar 3$Q$ states~\cite{Martin_PhysRevLett.101.156402,Akagi_JPSJ.79.083711,hayami_PhysRevB.91.075104}. 

\subsection{Edge States}
\label{sec:Edge Dispersion}

\begin{figure}[hbt!]
\begin{center}
\includegraphics[width=1.0 \hsize]{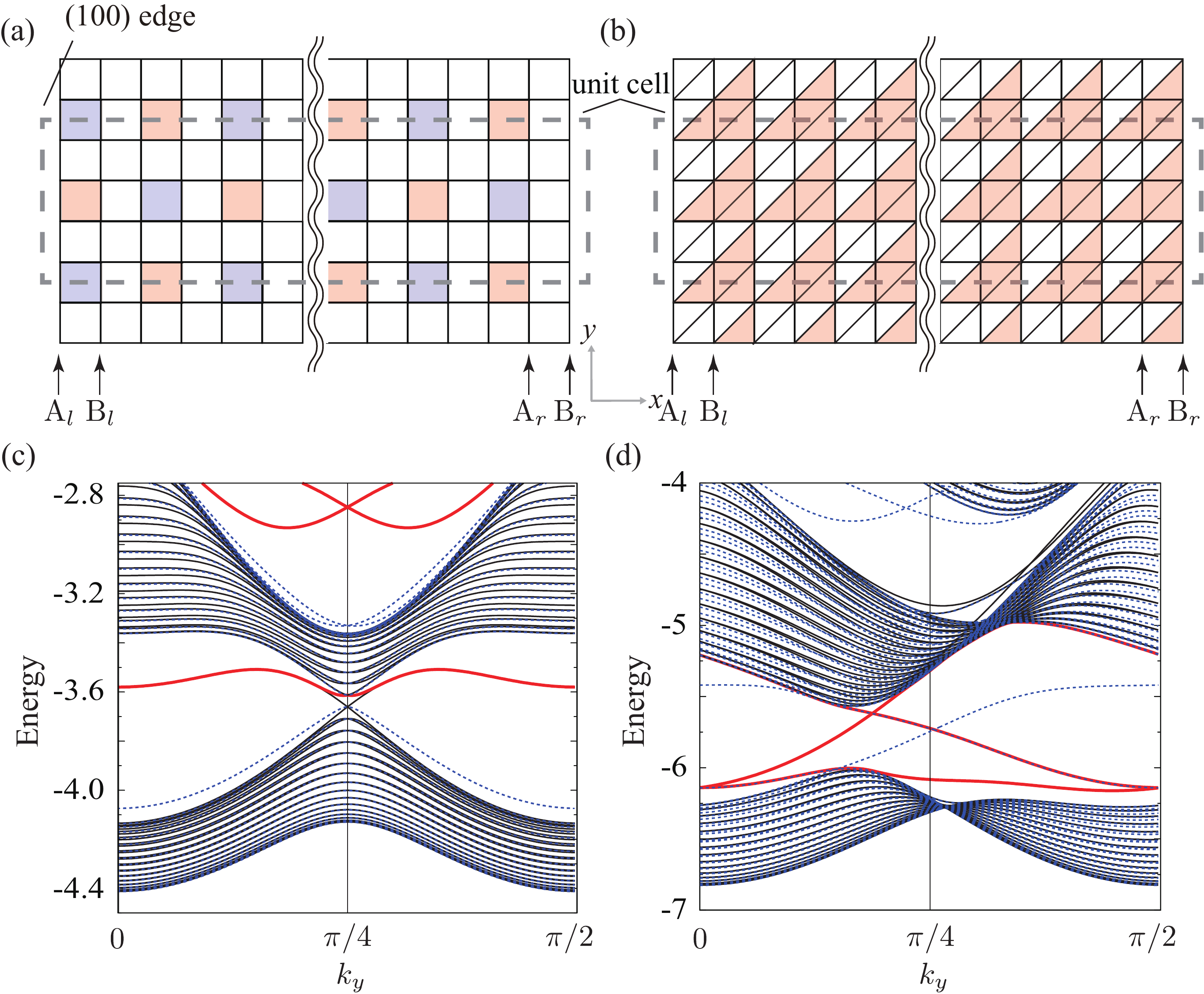} 
\caption{
\label{Fig:edge}
(Color online)
Schematic picture of the system with the (100) edges for (a) square lattice and (b) triangular lattice. 
The dashed boxes represent unit cells used for the calculations of the edge states; the unit cell contains 256 or 260 sites depending on where the edges are terminated. 
The red (blue) plaquettes represent positive (negative) values of (a) vector and (b) scalar chirality. 
(c), (d) Energy dispersions near the Fermi level at $n=0.125$ for (c) the 2$Q$ and (d) 3$Q$ UUDD states for the systems with open edges shown in (a) and (b), respectively. 
In (c)[(d)], the solid and dashed lines represent the dispersion in the systems with the (A$_l$, A$_r$) [(A$_l$, A$_r$)] and (B$_l$, B$_r$) [(A$_l$, B$_r$)] edges, while the thick red lines the edge states in the systems with the (A$_l$, A$_r$) [(A$_l$, A$_r$)]. 
}
\end{center}
\end{figure}

Reflecting the topologically nontrivial nature induced by ChDW in the multiple-$Q$ UUDD states, peculiar edge states are observed, as demonstrated in Fig.~\ref{Fig:edge}. 
Here, we consider the systems with the (100) edges: the 2$Q$ UUDD state on the square lattice [Fig.~\ref{Fig:edge}(a); see also Fig.~\ref{Fig:multipleQ_ponti}(e)] and the 3$Q$ UUDD state on the triangular lattice [Fig.~\ref{Fig:edge}(b); see also Fig,~\ref{Fig:multipleQ_ponti}(h)]. 
We adopt the periodic boundary condition in the (010) direction. 
There are four choices for the edges depending on where we cut the ChDW (two for each edge): 
${\rm A}_l$ or ${\rm B}_l$ for the left edge and ${\rm A}_r$ or ${\rm B}_r$ for the right edge, as shown in the figures. 
The edge states appear in the electronic structure in a different form depending on the choices, as demonstrated below. 

Figure~\ref{Fig:edge}(c) shows the band dispersions near the Fermi level at $n=0.125$ for the 2$Q$ UUDD state with antiferroic ChDW on the square lattice with the (100) edges. 
In this case, the edge states, which are represented by the thick red lines, appear around the Dirac nodes when we take the (A$_l$, A$_r$) edges. 
This is presumably owing to the nonzero vector chirality in the plaquettes on the edges. 
In fact, such edge states do not appear for the (B$_l$, B$_r$) edges, where the vector chirality vanishes in the plaquettes on the edges. 
We note that the two edge states are doubly degenerate each in this case. 
Meanwhile, for the (A$_l$, B$_r$) or (B$_l$, A$_r$) edges, one of the two edge states shows a similar dispersion to that in the (A$_l$, A$_r$) edge represented by thick red lines in Fig.~\ref{Fig:edge}(c), while the other edge state is similar to that in the (B$_l$, B$_r$) edge (not shown here). 

On the other hand, the 3$Q$ UUDD state on the triangular lattice shows gapless chiral edge states traversing the energy gap of the Chern insulator, irrespective of the choice of the edges, as shown in Fig.~\ref{Fig:edge}(d).  
These are topologically-protected edge states, as the partially ferroic ChDW is a Chern insulator with nonzero net component of the scalar chirality. 
Even in this situation, however, the chiral edge states behave differently depending on the choice of the edges. 
For instance, as shown in Fig.~\ref{Fig:edge}(d), when we change the right edge from A$_r$ to B$_r$ with keeping the left edge A$_l$, the chiral edge dispersion with a positive slope shows a drastic change. 
The result indicates that we can control the edge currents by the location of the edges, namely, by the phase of ChDW. 
Such phase-dependent edge states suggest a new possibility of controlling the electronic structures and transport properties by nanostructure of ChDW.

\section{Summary and Concluding Remarks} 
\label{sec:Conclusion}
To summarize, we have investigated the possibility of vector and scalar ChDW in itinerant magnets, focusing on the construction from multiple-$Q$ superpositions of the UUDD collinear spin structures. 
We have examined the stability of the multiple-$Q$ UUDD states in the Kondo lattice model with classical localized moments on square and triangular lattices, using three complementary methods: perturbative analysis, variational calculations, and Langevin dynamics simulations. 
Contrary to the common belief that the RKKY interaction stabilizes a helical state, all the results consistently indicate that the itinerant systems exhibit the multiple-$Q$ UUDD states in the limit of weak spin-charge coupling. 
This occurs when the Fermi surface is connected by the commensurate ordering vectors that are related with each other by rotational symmetry compatible to the lattice structure. 
Although they share the stabilization mechanism with the previously studied multiple-$Q$ states, we showed that the multiple-$Q$ UUDD states have greater flexibility; for instance, they can accommodate two-dimensional textures of vector and scalar chirality. 
We also found that ChDW associated with the multiple-$Q$ UUDD states bring about nontrivial topology in the electronic structures, such as massless Dirac semimetals and Chern insulators.  
In addition, we clarified that, reflecting the spatial modulation of vector and scalar chirality, the peculiar edge states appear in the topologically nontrivial states, which depend on how the ChDW are terminated at the edges. 
The results suggest the controllability of edge currents by the phase of ChDW. 

Finally, let us comment on the competition between the multiple-$Q$ UUDD states and multiple-$Q$ helical states with one-dimensional stripy ChDW~\cite{Solenov_PhysRevLett.108.096403,ozawa2015vortex}
In the current study, we found that the multiple-$Q$ UUDD state is more stabilized than the multiple-$Q$ helical ones with stripy ChDW by KPM-LD numerical simulations for $\mathbf{Q}_1=(\pi/2,\pi)$. 
We also confirmed that the situation is also similar for $\mathbf{Q}_1=(\pi/2,0)$ by the variational calculations (not shown here). 
Such preference is restricted to particular ordering vectors, $\mathbf{Q}_1=(\pi/2,0)$ or $\mathbf{Q}_1=(\pi/2,\pi)$; for other commensurate wave vectors, e.g., $\mathbf{Q}_1=(\pi/l,0)$ ($l$ is an integer larger than two), we need higher harmonics to constitute collinear orders, while we do not for helical ones. 
Let us take an example of $\mathbf{Q}_1=(\pi/3,0)$.
In order to construct a UUUDDD collinear state, one needs an additional ordering vector, $\mathbf{Q}_1'=(\pi,0)$, in addition to $\mathbf{Q}_1=(\pi/3,0)$, which leads to an energy cost even at the RKKY level. 
Thus, the multiple-$Q$ helical states with stripy ChDW will be more stable than the multiple-$Q$ UUUDDD states, at least, in the weak $J$ limit. 
The situation might be turned over when considering large $J$ and taking into account other contributions, such as the spin anisotropy and spin-orbit coupling. 
Such extensions are left for future study.   

\begin{acknowledgments}
The authors thank H. Kawamura for enlightening discussions in the early stage of the present study.  
The authors also thank K. Barros for the fruitful discussions on the KPM-LD simulations. 
The KPM-LD simulations were carried out at the Supercomputer Center, 
Institute for Solid State Physics, University of Tokyo.
R.O. is supported by the Japan Society for the Promotion of Science 
through a research fellowship for young scientists and the Program for Leading Graduate Schools (ALPS). 
This research was supported by KAKENHI (No.~24340076), the Strategic Programs for Innovative Research (SPIRE), MEXT, and the Computational Materials Science Initiative (CMSI), Japan. 
\end{acknowledgments}

\bibliographystyle{apsrev}
\bibliography{ref}

\end{document}